# Monte Carlo Study of Temperature-dependent Non-diffusive Thermal Transport in Si Nanowires

Lei Ma [1, 2], Riguo Mei[1, 2], Mengmeng Liu[3], Xuxin Zhao[2], Qixing Wu[2], Hongyuan Sun[2, *]

[1] Key Laboratory of Optoelectronic Devices and Systems of Ministry of Education and Guangdong Province, Shenzhen University, 518060, People's Republic of China

[2] Key Laboratory of New Lithium-ion Batteries and Mesoporous Materials of Shenzhen City, College of Chemistry and Environmental Engineering, Shenzhen University, 518060, People's Republic of China

[3] Department of Radiology, University of California, 94107, United States

*Corresponding author, Email: hysun163@163.com

ABSTRACT

Non-diffusive thermal transport has gained extensive research interest recently due to its important implications on fundamental understanding of material's phonon mean free path distributions and many nanoscale energy applications. In this work, we systematically investigate the role of boundary scattering and nanowire length on the non-diffusive thermal transport in thin silicon nanowires by rigorously solving the phonon Boltzmann transport equation (BTE) using a variance reduced Monte Carlo technique across a range of temperatures. The simulations use the complete phonon dispersion and spectral lifetime data obtained from first-principle density function theory calculations as input without any adjustable parameters. Our BTE simulation results show that the nanowire length plays an important role in determining the thermal conductivity of silicon nanowires. In addition, our simulation results suggest significant phonon confinement effect for the previously measured silicon nanowires. These findings are important for a comprehensive understanding of microscopic non-diffusive thermal transport in silicon nanowires.

**Introduction**

Nanowires, due to their unique electrical, thermal, and optical properties, have been a topic of extensive research interest for the past two decades[1–3]. Like other low-dimensional material structures[4,5], nanowires are promising for many practical applications, including, but not limited to, thermal management in nanoelectronic and optoelectronic devices[6–8], thermal interface materials[9,10], and thermoelectric energy conversion[11–14]. In particular, thermal transport in silicon nanowires (SNWs) have been extensively studied both experimentally and theoretically. For example, Volz and Chen computed the SNW thermal conductivity using molecular dynamics (MD) simulations and found nearly two orders of magnitude reduction in thin SNW thermal conductivity at room temperature[1]. In 2003, Li *et al.*[15] first experimentally measured the thermal conductivity of single-crystalline thin SNWs and observed substantial reduction in the measured thermal conductivity across a wide range of temperatures. They attributed the thermal conductivity reduction to increased phonon boundary scattering at the lateral walls of the nanowires and possible phonon confinement effect that may alter the bulk phonon dispersion relation. Two subsequent experimental studies[11,12] observed more than two orders of magnitude thermal conductivity reduction in their respectively synthesized on rough SNWs and SNW array with almost unaltered electrical conductivity and Seebeck coefficient, making them promising for high-performance and scalable thermoelectric materials.

On the other hand, various theoretical and numerical studies typically invoked MD simulations[16–19] and phonon Boltzmann transport equation (BTE)[2,20] to examine the role of boundary scattering, phonon confinement[21], phonon-surface-roughness scattering[22], and isotope scattering[16] on the thermal transport in SNWs. In particular, Chen *et al.*[20] used

Monte Carlo simulations to investigate the effect of boundary scattering and phonon confinement on the thermal conductivity of thin SNWs. Yang et al.[16] observed exponential reduction in SNW thermal conductivity by isotopic defects at room temperature. Martin et al.[23] studied the effect of surface roughness on the thermal conductivity of SNWs and found a quadratic dependence of the thermal conductivity on the ratio of diameter to surface roughness for sub-100nm sized nanowires. These simulation studies have helped gain insight into the nature of microscopic thermal transport in SNWs. However, these theoretical efforts typically neglect the impact of finite nanowire length on the nanowire thermal conductivity. Size effects can also occur when the nanowire length becomes comparable with the phonon mean free paths (MFPs). Non-diffusive thermal transport caused by finite nanowire length was only recently reported from experimental measurements on SiGe NWs with variable lengths[24]. In particular, Hsiao et al.[24] observed that ballistic thermal transport in micro-fabricated SiGe nanowires can persist over ~ 8.3 um, much longer than previously believed.

Although non-diffusive thermal transport in NWs plays an important role in many applications as the diameter and length of the nanowires incorporated in devices both become comparable with the heat-carrying phonon MFPs[25–31], to the best of our knowledge, few studies have devoted to a comprehensive theoretical and numerical understanding of non-diffusive thermal transport caused by both lengthscales simultaneously in such nanostructures. In this work, we systematically investigate the role of phonon boundary scattering and finite nanowire length on the non-diffusive thermal transport in SNWs. A common practice to measure the degree of non-diffusive thermal transport is the use of effective thermal conductivity for the given nanostructure. We calculate the effective thermal conductivity for SNWs of variable diameters, lengths, and boundary specularities across a range of temperatures by rigorously solving the full phonon BTE under the relaxation time approximation. The numerous simulations in this study use a recently developed variance reduced Monte Carlo technique to achieve high calculation accuracy and acceptable computational cost. The complete phonon dispersion and lifetime data calculated from first-principle density functional theory are used as inputs in the simulations without any adjustable parameters. We find that both the nanowire diameter and length play an important role in determining the thermal conductivity and interpretation for nanowire thermal conductivity must take into account the nanowire length when it becomes comparable with phonon MFPs. Our simulation results also suggest significant phonon confinement effect in the previously experimentally studied SNWs.

**Simulation Details**

To gain insight into how different lengthscales affect the thermal transport in thin NWs, the full spectral phonon BTE must be solved[32]. Since it is computationally expensive to solve the BTE using a deterministic approach[33,34], we resort to a recently developed deviational MC technique to rigorously solve the multidimensional phonon BTE to study the thermal transport physics in square SNWs[35–37]. Under the relaxation time approximation, the deviational spectral energy-based phonon BTE in its general form is given by[32]:

$$\frac{\partial e_\omega}{\partial t} + \vec{V_\omega} \cdot \nabla e_\omega = -\frac{e_\omega - e_{\omega 0}}{\tau_\omega} \quad (1)$$

where $\omega$ is the phonon angular frequency, $V_\omega$ is the frequency-dependent phonon group velocity, $\tau_\omega$ is the spectral phonon relaxation time, $e_\omega = \hbar\omega(f_\omega - f_\omega^{eq})$ represents the deviational phonon energy distribution that is the product of the phonon energy $\hbar\omega$ and the deviational phonon distribution $(f_\omega - f_\omega^{eq})$, and $e_{\omega 0}$ represents the local equivalent equilibrium deviational energy distribution function. The deviational phonon distribution $(f_\omega - f_\omega^{eq})$ is the difference between the local phonon distribution function $f_\omega$ and the Bose-Einstein distribution function $f_\omega^{eq} = \frac{1}{\exp\left(\frac{\hbar\omega}{k_B T_{eq}}\right)+1}$ evaluated at a reference temperature $T_{eq}$. The advantage of solving the deviational energy-based phonon BTE instead of the original form is two-fold[35,36]. On one hand, it allows the energy conservation to be conserved automatically by conserving the number of computational phonon particles in the simulation domain. One the other hand, it allows simulations with high accuracy to be accomplished since we only simulate the deviation from a known reference equilibrium distribution that has analytical transport solution and therefore does not cause any stochastic noise in the final solution.

Details of the MC algorithm has been extensively described elsewhere[35,36,38–40] and are discussed briefly here. In this work, we study the thermal conductivity of SNWs across a set of temperatures where relaxation time approximation is applicable for silicon. For each temperature, a temperature difference $\Delta T = T_h - T_c = 1\ K$ is imposed on the two ends of the SNW, where $T_h$ is the hot end temperature and $T_c$ is the cold end temperature. By taking the reference temperature to be $T_c$, computational phonon particles are emitted from the hot end of the NW and they subsequently travel inside the NW subject to anharmonic scattering and boundary scattering. Phonon boundary scattering occurs if one computational particle encounters the lateral boundary of the nanowire during one specific advection step. In the occurrence of boundary scattering, resetting the phonon traveling direction depends upon the boundary specularity. Normally, a random number is drawn and compared to the boundary specularity. If the random number is smaller than the specularity, phonons are specularly reflected at the boundary; otherwise, it is diffusely scattered and its traveling direction is randomized into the simulation domain. Phonon-phonon scattering occurs if the particle does not collide with the boundaries in one advection step. All the phonon states, including frequency, group velocity, branch, and travelling direction, are reset in the event of phonon-phonon scattering. Since the temperature difference is small across the entire simulation domain, the scattering term in the phonon BTE can be linearized (as explained in ref. 34), resulting a significant simplification and efficiency gain because the computational particles can be simulated sequentially. The simulation of one particle ends when it is absorbed by either the hot end or the cold end. By looping the particles, we sample the contribution from all the simulated particles to find the heat flux and local temperature. The effective thermal conductivity of the NW is calculated through: $k_{\text{eff}} = Q/A\Delta T$, where Q is the heat flux across the NW and A is the cross-sectional area, respectively. In the limit of long nanowire, size effect results merely from the phonon boundary scattering. However, when the wire length becomes comparable with the phonon MFP, both the wire diameter and length affect the thermal transport[41].

**Results and Discussion**

The developed simulation framework is applied to single-crystalline silicon nanowires of variable diameters and lengths to study the non-diffusive thermal transport behavior across a range of temperatures and lengthscales. The simulation takes as input the full spectral phonon properties calculated by first-principle density functional theory and uses no adjustable parameters[42]. To ensure the accuracy of the simulations, we use at least five million computational particles for the smallest diameter case and more particles for nanowires with larger diameters.

Figure 1 shows the axial temperature distribution versus three different nanowire lengths for each of the three nanowires with distinct diameters at room temperature (300K). The results are shown for silicon nanowires with diffuse scattering at the boundaries of the nanowires. The axial temperature is averaged over the nanowire cross-sectional area and is represented relative to the reference temperature (i.e. the cold end temperature). For nanowire with a diameter of 22 nm, as shown in Fig. 1(a), the temperature distribution is nearly linear along the axial direction when the nanowire length is larger than 1 sum. In addition, no significant temperature jump occurs at the two ends of the nanowire for very long nanowires. The absence of temperature jumps at the nanowire ends for long nanowires in Fig. 1(a) simply indicates that no non-diffusive transport effect is induced by the finite nanowire length[43] and we can use Fourier's law to describe the thermal transport in the nanowire with a modified effective thermal conductivity since the size effect is caused only by the finite nanowire diameter[41,44–46]. However, when the nanowire length is less than 200 nm, we observe curved temperature distribution and significant temperature jumps at the boundaries and the magnitude of the temperature increases with decreasing nanowire length. The temperature jump is a clear indication of non-diffusive thermal transport caused by the finite length of the nanowire in addition to the size effect caused by the finite diameter. For nanowires of larger diameters, as shown in Fig. 1(b) and 1(c), the axial temperature profile is still close to linear. However, as the nanowire diameter increases, the magnitude of the temperature jump increases consistently and the onset of non-diffusive thermal transport caused by the finite nanowire length shifts to larger lengths. This is so because the effective phonon MFPs become longer in a larger nanowire[41].

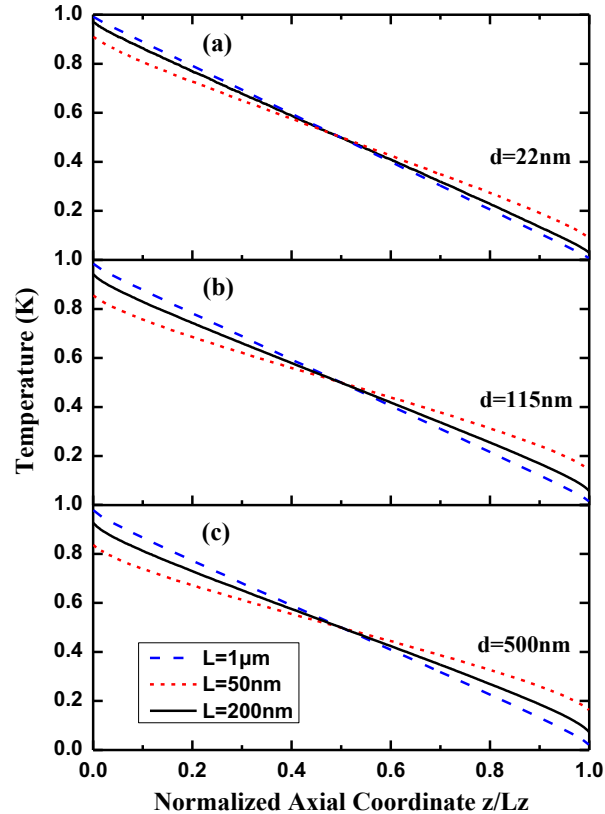

Figure 1. Average axial temperature profile for nanowires of three different diameters at three distinct nanowire lengths: (a) diameter = 22 nm, (b) diameter = 115 nm, (c) diameter = 500 nm.

As discussed before, Li *et al.*[15] previously measured the diameter-dependent thermal conductivities of silicon nanowires of several micrometers in length across a wide range of temperatures using a micro-fabricated measurement setup. We compare our calculated size-dependent thermal conductivities with their measurement results in Fig. 2 with curves representing simulation data and dots representing experimental results. The simulation results are obtained for diffuse-boundary-scattering nanowires with a nanowire length corresponding to the experimental values. Note that our simulation only covers the temperature range from 100 K to 350 K since we calculate the nanowire thermal conductivity by solving the phonon BTE under the relaxation time approximation that is not valid for temperatures below 100 K in crystalline silicon. As shown in Fig. 2, the simulation results closely follow the same trend as the experimental data. However, for all the simulated nanowires, the calculated thermal conductivities are typically higher than the measurement results. Since we use first-principle calculated properties in the simulations, this discrepancy largely indicates that strong phonon confinement effect occurs for small nanowires that flattens the phonon dispersion relation and therefore reduces the phonon group velocities, resulting in much lower measured nanowire thermal conductivities. The significantly lower measured thermal conductivities might also be attributable to the rough boundary scattering, as shown in another previous experimental effort[12,13].

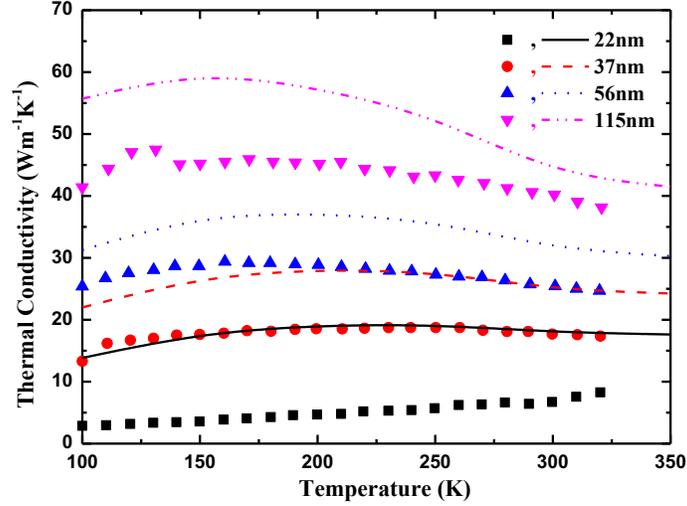

Figure 2. Comparison of computed silicon nanowire effective thermal conductivities with experimentally measured results across a range of diameters and temperatures.

Non-diffusive thermal transport in SiGe nanowires with diameter on the order of 100 nm was observed by Hsiao. *et al.*[24] for nanowire length as long as ~ 8.3 um at room temperature. To examine the non-diffusive transport effect caused by the finite nanowire length, we carry out BTE simulations for a set of nanowire lengths for different nanowire diameters (22 nm, 115 nm, and 500 nm) at different temperatures. Figure 3 shows the computed effective thermal conductivities as a function of length for three diffuse-boundary-scattering nanowires of different diameters at three different temperatures, 100 K, 200 K, and 300 K. Clearly the thermal conductivity is significantly reduced when the nanowire length becomes comparable or shorter than the phonon MFP. This suggests that thermal conductivity measurements must take into account the effect of finite nanowire length if the nanowire is not sufficiently long to eliminate the size effect caused by its length. In general, as mentioned above, the onset of non-diffusive thermal transport induced by the finite nanowire length occurs at a larger length value for a bigger nanowire[41]. In addition, as we lower the temperature, the onset of non-diffusive thermal transport also slightly shifts to larger nanowire length since the phonon MFPs increase significantly with decreasing temperature. Here, we only present results for wires with diameters less than 1 um, but we note that for wires of diameter of the a few micrometers, the non-diffusive thermal transport can occur at much a larger wire length.

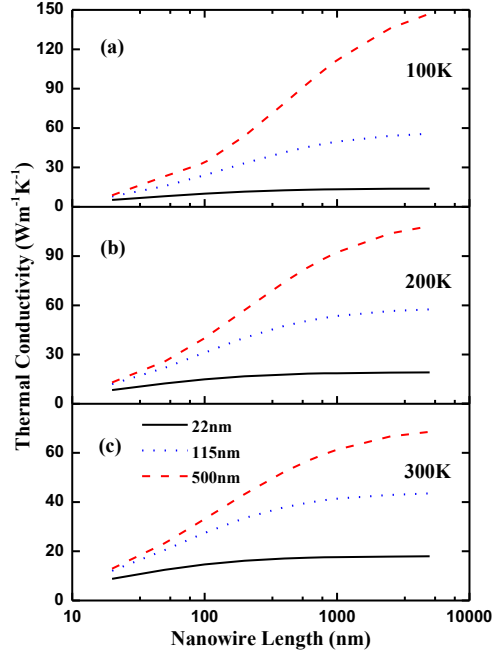

Figure 3. Computed effective thermal conductivities of silicon nanowires as a function of nanowire length for three different nanowire diameters (22 nm, 115 nm, and 500 nm) across three different temperatures: 100 K, 200 K, and 300 K.

Note that the direct observation of non-diffusive thermal transport persisting ~ 8.3 um in Hsiao *et al.*'s study is only possible with perfectly specular boundary reflection[24]. Here, we systematically examine the role of nanowire boundary specularity on the non-diffusive thermal transport in silicon nanowires[43]. The boundary specularity is a measure of the fraction of phonons that are specularly reflected at the nanowire boundary. Figure 4 shows the calculated effective thermal conductivities versus nanowire length for three nanowires with different diameters (56 nm, 115 nm, and 500 nm) and boundary specularities at room temperature (300K). It is clear that boundary specularity increases the effective thermal conductivity since it does not cause thermal resistance along the heat flux direction[43]. We also observe that increasing boundary specularity shifts the onset of non-diffusive thermal transport to larger nanowire length, meaning that it would be easier to observe non-diffusive thermal transport in a nanowire with specularly reflecting boundaries. We note that for nanowires with perfectly specular boundary reflection, any size effect would come from the finite nanowire length. For partially diffuse and partially specular nanowires, size effects would come from both the finite nanowire diameter and the finite nanowire length. Another interesting feature to observe is that the effect of boundary specularity is relatively weaker when the nanowire length is short. This occurs due to the fact that non-diffusive effect is dominated by the finite nanowire length for short nanowires rather than the diameter.

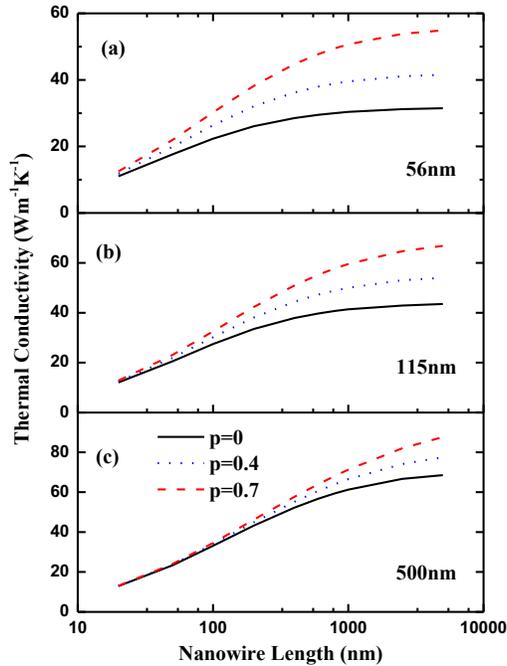

Figure 4. Computed effective thermal conductivities of silicon nanowires as a function of nanowire length across a set of nanowire boundary specularities. The boundary speculariry measures the fraction of boundary phonons that are specularly reflected.

**Conclusion**

In summary, we report the study of non-diffusive thermal transport in single-crystalline silicon nanowires by rigorously solving the full spectral phonon Boltzmann transport equation across a range of temperatures and lengthscales. We observe significant reduction in thermal conductivity for silicon nanowires of a few tens to a few hundreds of nanometers in diameter and obtain reasonably good agreement between our calculation results and previously reported experimental data. Our simulation results indicate the importance of accounting for the size effects caused by both the nanowire diameter and length in interpreting the experimental thermal conductivity measurement results on nanowires. The simulation results also imply that significant phonon confinement effect are present for the those previously measured silicon nanowires. These results help gain fundamental insight into the microscopic non-diffusive thermal transport in nanowire systems.

Author Contributions

L.M. and R.M contributed equally to this work.

Financial Interests

The authors declare no competing financial interests.

Acknowledgement

This work is supported by National Natural Science Foundation of China (No. 51306125) and by Shenzhen Research Foundation of Science & Technology (KQCX20140519105122378, JCYJ20130329113322731).

**References**

[1] S.G. Volz and G. Chen, Appl. Phys. Lett. **75**, 2056 (1999).

[2] R. Yang, G. Chen, and M.S. Dresselhaus, Nano Lett. **5**, 1111 (2005).

[3] R. Chen, A.I. Hochbaum, P. Murphy, J. Moore, P. Yang, and A. Majumdar, Phys. Rev. Lett. **101**, 105501 (2008).

[4] M.N. Luckyanova, J. Garg, K. Esfarjani, A. Jandl, M.T. Bulsara, A.J. Schmidt, A.J. Minnich, S. Chen, M.S. Dresselhaus, Z. Ren, E.A. Fitzgerald, and G. Chen, Science (80-. ). **338**, 936 (2012).

[5] R. Jia, L. Zeng, G. Chen, and E.A. Fitzgerald, arXiv preprint arXiv: 1610.02102 (2016).

[6] Y. Cui, Z. Zhong, D. Wang, W.U. Wang, and C.M. Lieber, Nano Lett. **3**, 149 (2003).

[7] F. Schäffler, W. Lu, and C.M. Lieber, J. Phys. D. Appl. Phys. **39**, R387 (2006).

[8] B. Tian, X. Zheng, T.J. Kempa, Y. Fang, N. Yu, G. Yu, J. Huang, and C.M. Lieber, Nature **449**, 885 (2007).

[9] A.M. Marconnet and K.E. Goodson, J. Heat Transfer **135**, 061601 (2013).

[10] A.M. Marconnet, M.A. Panzer, and K.E. Goodson, Rev. Mod. Phys. **85**, 1295 (2013).

[11] A.I. Boukai, Y. Bunimovich, J. Tahir-Kheli, J.-K. Yu, W. a Goddard, and J.R. Heath, Nature **451**, 168 (2008).

[12] A.I. Hochbaum, R. Chen, R.D. Delgado, W. Liang, E.C. Garnett, M. Najarian, A. Majumdar, and P. Yang, Nature **451**, 163 (2008).

[13] J.P. Feser, J.S. Sadhu, B.P. Azeredo, K.H. Hsu, J. Ma, J. Kim, M. Seong, N.X. Fang, X. Li, P.M. Ferreira, S. Sinha, D.G. Cahill, J.P. Feser, J.S. Sadhu, B.P. Azeredo, K.H. Hsu, and J. Ma, J. Appl. Phys. **112**, 114306 (2012).

[14] R. He, D. Kraemer, J. Mao, L. Zeng, Q. Jie, Y. Lan, C. Li, J. Shuai, H.S. Kim, Y. Liu, D. Broido, C.-W. Chu, G. Chen, and Z. Ren, Proc. Natl. Acad. Sci. **113**, 13576 (2016).

[15] D. Li, Y. Wu, P. Kim, L. Shi, P. Yang, A. Majumdar, and D. Li, Appl. Phys. Lett. **83**, 2934 (2003).

[16] I.S. Nanowires, N. Yang, G. Zhang, and B. Li, Nano Lett. **8**, 276 (2008).

[17] N. Yang, G. Zhang, and B. Li, Nano Today **5**, 85 (2010).

[18] X. Li, K. Maute, M.L. Dunn, and R. Yang, Phys. Rev. B **81**, 245318 (2010).

[19] Q. Liao, L. Zeng, Z. Liu, and W. Liu, Sci. Rep. **6**, 34999 (2016).

[20] Y. Chen, D. Li, J.R. Lukes, and A. Majumdar, J. Heat Transfer **127**, 1129 (2005).

[21] I. Ponomareva, Nano Lett. **7**, 1155 (2007).

[22] L. Liu, X. Chen, L. Liu, and X. Chen, J. Appl. Phys. **107**, 033501 (2010).

[23] P. Martin, Z. Aksamija, E. Pop, and U. Ravaioli, Phys. Rev. Lett. **102**, 125503 (2009).

[24] T. Hsiao, H. Chang, S. Liou, M. Chu, S. Lee, and C. Chang, Nat. Nanotechnol. **8**, 534 (2013).

[25] A.J. Minnich, J.A. Johnson, A.J. Schmidt, K. Esfarjani, M.S. Dresselhaus, K.A. Nelson, and G. Chen, Phys. Rev. Lett. **107**, 095901 (2011).

[26] J.A. Johnson, A.A. Maznev, J. Cuffe, J.K. Eliason, A.J. Minnich, T. Kehoe, C.M.S. Torres, G. Chen, and K.A. Nelson, Phys. Rev. Lett. **110**, 025901 (2013).

[27] K.M. Hoogeboom-Pot, J.N. Hernandez-Charpak, X. Gu, T.D. Frazer, E.H. Anderson, W. Chao, R.W. Falcone, R. Yang, M.M. Murnane, H.C. Kapteyn, and D. Nardi, Proc. Natl. Acad. Sci. **112**, 4846 (2015).

[28] Y. Hu, L. Zeng, A.J. Minnich, M.S. Dresselhaus, and G. Chen, Nat. Nanotechnol. **10**, 701 (2015).

[29] L. Zeng, K.C. Collins, Y. Hu, M.N. Luckyanova, A.A. Maznev, S. Huberman, V. Chiloyan, J. Zhou, X. Huang, K.A. Nelson, and G. Chen, Sci. Rep. **5**, 17131 (2015).

[30] V. Chiloyan, L. Zeng, S. Huberman, A.A. Maznev, K.A. Nelson, and G. Chen, Phys. Rev. B **93**, 155201 (2016).

[31] V. Chiloyan, L. Zeng, S. Huberman, A.A. Maznev, K.A. Nelson, and G. Chen, J. Appl. Phys. **120**, 025103 (2016).

[32] G. Chen, *Nanoscale Energy Transport and Conversion* (Oxford University Press, New York, 2005).

[33] R. Yang, G. Chen, M. Laroche, and Y. Taur, J. Heat Transfer **127**, 298 (2005).

[34] L. Zeng and G. Chen, J. Appl. Phys. **116**, 064307 (2014).

[35] J.-P.M. Péraud and N.G. Hadjiconstantinou, Phys. Rev. B **84**, 205331 (2011).

[36] J.-P.M. Péraud and N.G. Hadjiconstantinou, Appl. Phys. Lett. **101**, 153114 (2012).

[37] J.-P.M. Péraud, C.D. Landon, and N.G. Hadjiconstantinou, Annu. Rev. Heat Transf. **17**, 205 (2014).

[38] Q. Hao, G. Chen, and M.-S. Jeng, J. Appl. Phys. **106**, 114321 (2009).

[39] L. Zeng, Experimental and Numerical Investigation of Phonon Mean Free Path Distribution, Massachusetts Institute of Technology, 2013.

[40] L. Zeng, Studying Phonon Mean Free Paths at the Nanoscale: Modeling and Experiments, Massachusetts Institute of Technology, 2016.

[41] L. Zeng, V. Chiloyan, S. Huberman, A.A. Maznev, J.M. Peraud, G. Nicolas, K.A. Nelson, and G. Chen, Appl. Phys. Lett. **108**, 063107 (2016).

[42] K. Esfarjani, G. Chen, and H.T. Stokes, Phys. Rev. B **84**, 085204 (2011).

[43] G. Chen, Phys. Rev. B **57**, 14958 (1998).

[44] K. Fuchs and N.F. Mott, Math. Proc. Cambridge Philos. Soc. **34**, 100 (1938).

[45] E.H. Sondheimer, *The Mean Free Path of Electrons in Metals* (Taylor and Francis, Oxford, 1952).

[46] A. Vega-Flick, R.A. Duncan, J.K. Eliason, J. Cuffe, J.A. Johnson, J.-P.M. Peraud, L. Zeng, Z. Lu, A.A. Maznev, E.N. Wang, J.J. Alvarado-Gil, M. Sledzinska, C. Sotomayor-Torres, G. Chen, and K.A. Nelson, AIP Adv. **6**, 121903 (2016).